\documentclass[preprint]{aastex}

\usepackage{graphicx}
\usepackage{natbib}

\newcommand{\kepler}{{\sl Kepler}}
\newcommand{\ra}{\alpha_{2000}}
\newcommand{\dec}{\delta_{2000}}

\newcommand{\hms}[3]{#1^\mathrm{h}\,#2^\mathrm{m}\,#3^\mathrm{s}}
\newcommand{\dms}[3]{#1^\mathrm{\circ}\,#2'\,#3''}

\begin{document}

\title{Stellar statistics along the ecliptic and the impact on the K2 mission concept}

\author{Andrej Pr\v sa}
\affil{Villanova University, Dept.~of Astrophysics and Planetary Science, 800 E Lancaster Ave, Villanova, PA 19085}
\affil{Jeremiah Horrocks Institute, University of Central Lancashire, Preston, PR1 2HE}
\email{aprsa@villanova.edu}

\and

\author{Annie Robin}
\affil{Institute Utinam, CNRS UMR6213, Universit\'{e} de Franche-Comt\'{e}, OSU 
THETA de Franche-Comt\'{e}-Bourgogne, Besan\c con, France}

\and

\author{Thomas Barclay}
\affil{NASA Ames Research Center, M/S 244-30, Moffett Field, CA 94035, USA}
\affil{Bay Area Environmental Research Institute, 596 1st Street West, Sonoma, CA 95476}

\begin{abstract}
K2 is the mission concept for a repurposed \kepler~mission that uses two reaction wheels to maintain the satellite attitude and provide $\sim$85 days of coverage for ten 105-deg${}^2$ fields along the ecliptic in the first 2.5 years of operation. We examine stellar populations based on the improved Besan\c con model of the Galaxy, comment on the general properties for the entire ecliptic plane, and provide stellar occurrence rates in the first 6 tentative K2 campaigns grouped by spectral type and luminosity class. For each campaign we distinguish between main sequence stars and giants and provide their density profile as a function of galactic latitude. We introduce the crowding metric that serves for optimized target selection across the campaigns. For all main sequence stars we compute the expected planetary occurrence rates for three planet sizes: 2-4\,$R_\oplus$, 4-8\,$R_\oplus$, and 8-32\,$R_\oplus$ with orbital periods up to 50 days. In conjunction with Gaia and the upcoming TESS and Plato missions, K2 will become a gold mine for stellar and planetary astrophysics.
\end{abstract}

\section{Introduction}

Space-borne missions CoRoT \citep{baglin2003} and \kepler~\citep{borucki2010} revolutionized stellar and planetary physics by providing us with ultra-precise photometric data of $\sim$30\,ppm. The duty cycle of observations exceeded 90\% and, for the first time, we had a nearly uninterrupted time coverage of over 200,000 objects. The two missions predominantly catered for two overlapping communities, the exoplanetary science and asteroseismology. The detection of extra-solar planets using the transit method boosted their numbers from dozens to nearly 5000 \citep{batalha2013, burke2014, rowe2014}, and the number is growing still as data are being mined. At the same time, asteroseismology witnessed an explosion in novel techniques and exciting new results \citep{chaplin2011}, ranging from main sequence B stars \citep{papics2013} to solar-like oscillations in red giants \citep{gaulme2013}. The overlap between the fields is provided by using asteroseismic techniques that provide fundamental properties of planet candidate host stars, which in turn enables exoplanet researchers to obtain precise fundamental properties of planets \citep{huber2013}. Unfortunately, CoRoT suffered from a computer failure in November 2012 and attempts to restore it ceased in June 2013. \kepler~lost a second reaction wheel in May 2013, causing the telescope to go to a prolonged safe-mode; attempts to bring Kepler back to operational state ceased in July 2013. However, this did not imply that \kepler~is retired; a proposal to use solar photon pressure to balance a 2-wheel \kepler~satellite enabled a continued operation. For this balancing to work, the telescope must point approximately in the direction of the ecliptic, so the observations of the initial \kepler~field are no longer possible. The spacecraft can hold pointing within $+50^\circ$ and $-30^\circ$ of its velocity vector in the orbital plane (approximately the ecliptic), with the two remaining reaction wheels holding the cross-boresight pointing steady. The spacecraft roll is minimized through regular thruster firing windows. The satellite can remain stable in roll for up to 85 days with a fuel budget that allows for a 2-3 year mission duration. A new mission concept, K2 \citep{howell2014}, builds on this engineering constraint. The science case arose from the community response to the whitepaper call for the repurposed mission and the concept was submitted to NASA HQ for the 2014 Senior Review (pending at the time of this writing). The first engineering observations utilizing the K2 mission design concept were obtained in October 2013 and the first full campaign-length test began in March 2014. This field lies in the direction near the galactic anti-center ($\ra = \hms{6}{33}{11.1}$, $\dec = \dms{21}{35}{16}$) and includes M35 and NGC 2158. The subsequent fields will be observed for 83 days; the duration is limited by solar illumination. If selected, K2 will observe upwards of 40,000-80,000 targets over the first year in 4 distinct fields.

\kepler~is a 0.95-m telescope with a 105-deg${}^2$ field of view. The early science commissioning run from October 2013 to February 2014 showed that the precision of K2 photometry for a $V=12$ star is $\sim$400\,ppm for the 30-min long cadence exposure and $\sim$80\,ppm for an integrated 6-hr exposure. The precision primarily depends on the spacecraft attitude jitter; the point spread function (PSF) of the K2 field is within 5\% of the original \kepler~field, and the degradation in precision due to a solar-induced drift is $\sim$4-fold \citep{howell2014}. Early science demonstration for WASP-28, a hot jupiter orbiting a Sun-like star in a 3.4-day orbit, corresponds to a 6-hr integrated noise level of $84$\,ppm.

This work employs the updated Besan\c con model of the Galaxy \citep[Robin et al.~2014, submitted]{robin2003} to simulate stellar populations along the ecliptic. The K2 mission has the potential observe $\sim$250,000 stars, and selecting targets hinges crucially on the representative population within each K2 field. The goal of this paper is to study the bulk properties and to serve as a guide to stellar populations along the ecliptic. This information can be used to better understand different populations from which the K2 targets are drawn, and to enable debiasing of any results that stem from statistical analyses of K2 campaigns.

\section{The new Besan\c con model}

The Besan\c con model of the Galaxy \citep[hereafter BGM]{robin2003} has served as one of the premier stellar population models. The model is by no means the only choice -- alternatives being TRILEGAL \citep{girardi2005}, the models of \citet{ng1997} and \citet{vallenari1999}, as well as the BGM-derived approach Galaxia \citep{sharma2011}. Here we use the scheme of the BGM described in \citet{robin2003}, updated by \citet{reyle2009} for the warp and flare parameters, by \citet{robin2012b} for the bulge and bar region, by Robin et al.~(2014; submitted) for the thick disc and halo shapes. The extinction model used is based on \citet{marshall2006}; for the regions not included in this study the BGM uses a thin disc of diffuse extinction \citep{robin2003}.

The model features four stellar populations: a thin disc, a bar, a thick disc and a spheroid. The star formation rate is constant over 10 Gyr for the thin disc population and is assumed to be a single burst in other populations with ages of 8, 12 and 14 Gyr, respectively. Each population has it own density law that were derived from wide survey data fits. Basel 3.1 \citep{westera1999} atmosphere models are used to compute the photometry in various systems (Johnson-Cousins, 2MASS, Spitzer, GALEX, UVIT).

This version of BGM does not take binary and multiple stars into account, but a new revision by \citet{czekaj2014} builds in binary populations as well, following the formalism of \citet{arenou2011}. For the purpose of this simulation only single stars are generated, so for remote populations the luminosity functions represent more systems than single stars. There are also no open clusters in the simulation.

\section{Stellar statistics}

\begin{figure}[t]
\includegraphics[width=\textwidth]{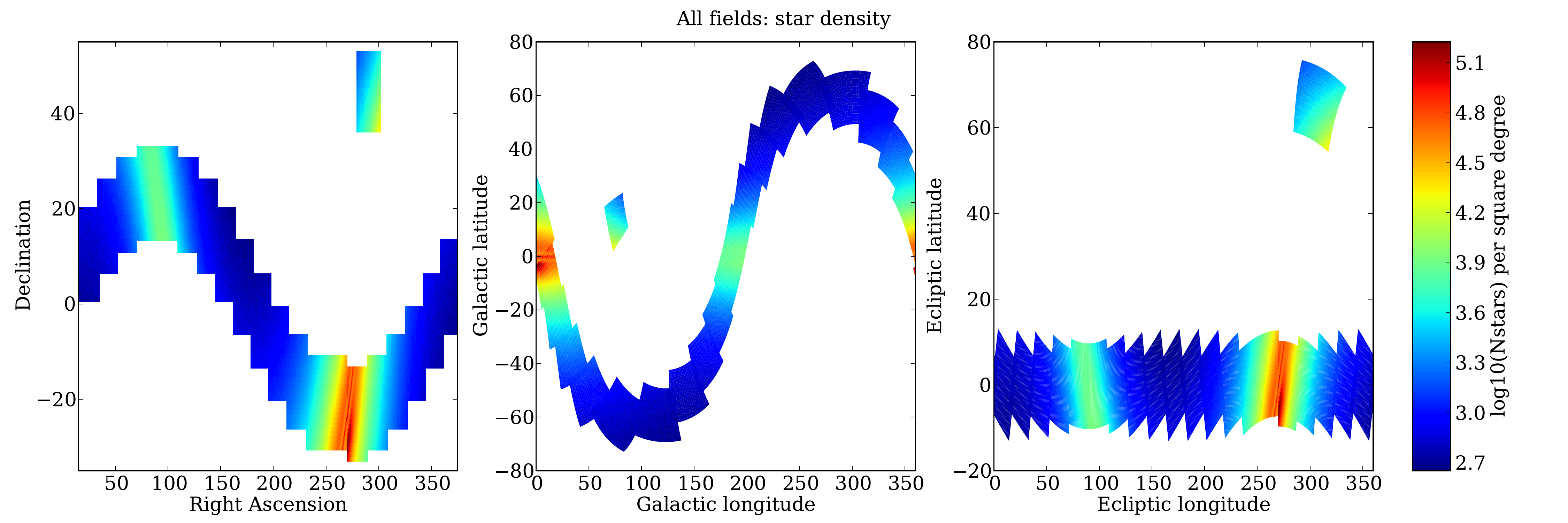} \\
\caption{
\label{fig:density}
Star counts (in log10 scale, $Kp =$7-17) in equatorial, galactic and ecliptic coordinates for 20 fields along the ecliptic. The counts are computed using the 2014 Besan\c con model of the Galaxy and range from $\sim$500 to $\sim$160,000 stars per deg${}^2$. The original \kepler~field is included for comparison purposes.
}
\end{figure}

The new BGM model has been run for 20 distinct fields along the ecliptic. The fields are $20^\circ \times 20^\circ$ in size and rectilinear in equatorial coordinates; the resolution element is 1-deg${}^2$. The fields are then transformed into galactic and ecliptic systems and star counts are performed for each 1-deg${}^2$ cell. Fig.~\ref{fig:density} depicts the number of stars in each cell for all 20 fields, and the original \kepler~field. The simulated stars range from $Kp = 7$--17 and cover all spectral types and luminosity classes in this magnitude range. All parameters of the model (used evolutionary tracks, atmosphere models, age-metallicity and age-velocity relations, the extinction model and radial scale length) can be found in \citet{czekaj2014}.

\begin{figure}[t!]
\includegraphics[width=\textwidth]{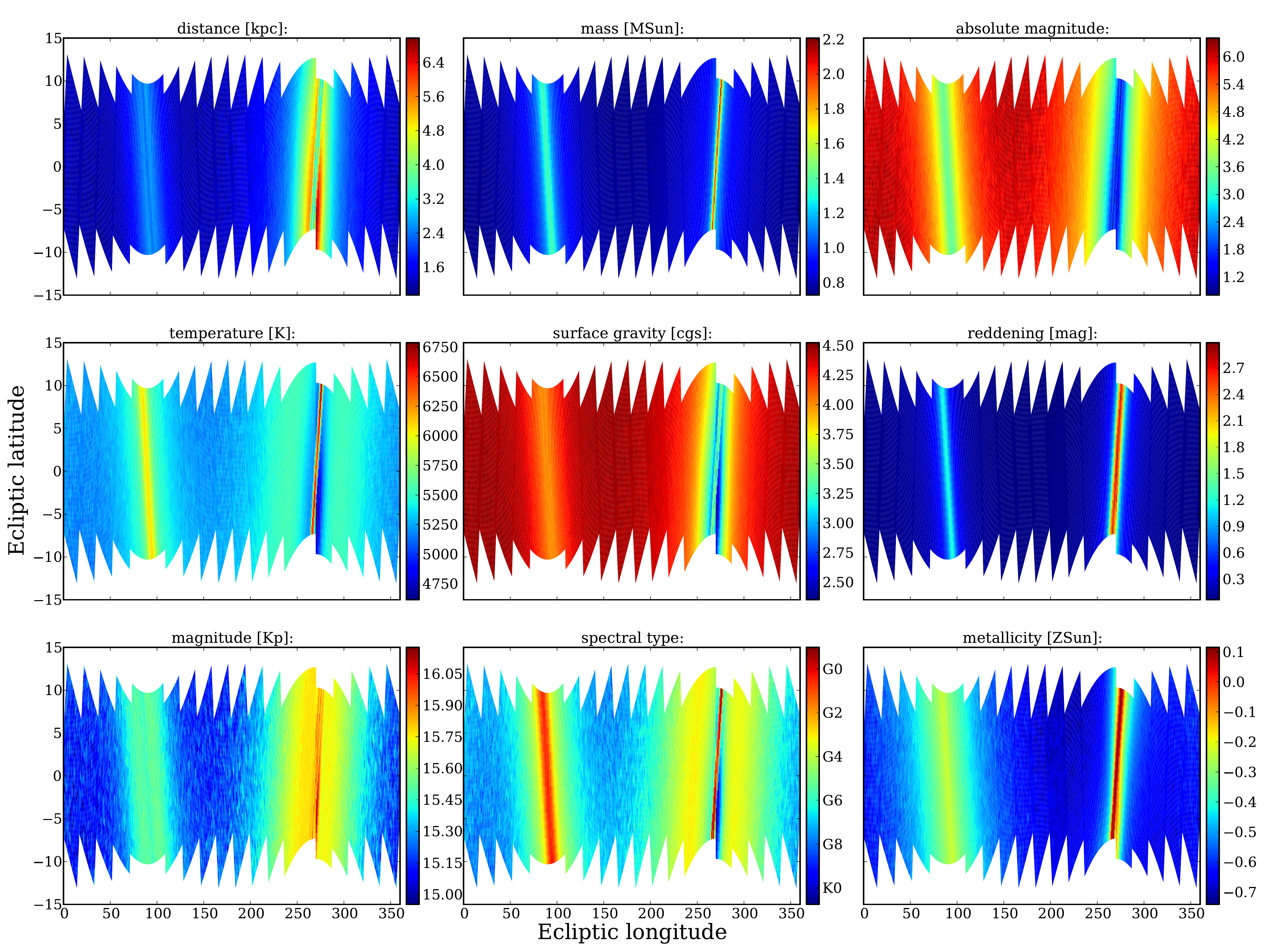} \\
\caption{
\label{fig:stellarpops}
Distributions of principal stellar parameters averaged over 1-deg${}^2$ in the $Kp=$7-17 range. All plots are given in ecliptic coordinates. Color bars denote the corresponding parameter values. Note that the values do not span the whole range of stellar parameters, only the \emph{average} values per 1-deg${}^2$. 
}
\end{figure}

It is instructive to review the predictions of BGM for single star populations. Fig.~\ref{fig:stellarpops} depicts 9 principal parameters of the stellar populations along the ecliptic in the $Kp=$7-17 magnitude range. Below we provide a brief commentary on each.

\begin{description}
\item[Distances.] The distance to the targets along the ecliptic is up to 2\,kpc, except close to the galactic plane, where it climbs above $\sim$3\,kpc. Baade's window, the largest transparent area towards the galactic bulge ($\ra = \hms{18}{03}{36}$, $\dec = \dms{-30}{02}{00}$), corresponds to the largest distances, upwards of 6\,kpc, where we can see the old, evolved population of stars.

\item[Masses.] In the selected magnitude range, the galactic plane hosts an intrinsically more massive population, averaging to a mass slightly larger than the Sun, and the bulge is responsible for the peak of about $2.2\,M_\odot$. Higher latitudes, on the other hand, are dominated by low-mass stars where the masses average to slightly less than the Sun.

\item[Absolute magnitudes.] The more massive population in the galactic plane corresponds to intrinsically brighter stars, whereas higher galactic latitudes are, on average, dominated by fainter stars. The values correspond to the \emph{average} absolute magnitude across the 1-deg${}^2$ cell that is magnitude-selected, which is why the values are dominated by bright objects (the Malmquist bias).

\item[Effective temperatures.] The mean effective temperature in the galactic disk is dominated by a younger, hotter population, whereas the regions towards galactic poles are notably cooler, dominated by the late-type field stars.

\item[Surface gravity.] Given in $cgs$ units ($\log g$), high surface gravity (dwarfs) dominate the higher latitudes of the galaxy while low surface gravity (subgiants and giants) dominate the disk, and most notably the bulge.

\item[Interstellar extinction.] The dust obscuration in the galactic plane is the main driver for the peak in extinction that averages to $\sim$2.5 magnitudes (with large variations, of course: some local patches are completely obscured and others, such as Baade's window, are mostly transparent), and drops to substantially smaller (but not completely negligible) levels farther from the galactic plane.

\item[Apparent magnitudes.] The simulated sample spans the $Kp =$7-17 magnitude range. In general, stellar population and interstellar extinction properties shape the distribution of magnitudes along the ecliptic. In the disk, obscuration wins out and results in a fainter sample; intermediate and higher latitudes are thus brighter. The region towards the galactic center is again on the faint end, partly because of obscuration and partly because of the distance to the bulge.

\item[Spectral types.] The bulk of \kepler~targets are in the $FGK$ range, with earlier types found in the disk of the galaxy and later types at intermediate and high latitudes. The effect of averaging across 1-deg${}^2$ is particularly notable here, rendering the span of average spectral types essentially across the $G$ range. The striking disparity near Baade's window is due to distinct populations -- old, evolved population in the bulge and young, hot population in the disk.

\item[Abundances.] When compared to the Sun, most of the Galaxy is on average under-abundant in metals, except for the bulge and the flaring thin disk. This is clearly evident in the abundance distribution, ranging from 20\% under-abundant ($[M/H] = -0.7$), typical of the thick disk population, to 25\% over-abundant ($[M/H] = +0.1$) towards the bulge.

\end{description}

With all this demonstrated variety along the ecliptic, the K2 mission is particularly well suited for population studies. Selected campaigns will probe inherently different populations and, for the first time, provide a nearly uninterrupted photometric coverage of tens of thousands of stars per field that will enable critical comparisons with, and calibrations of, the theoretical stellar population models.

\subsection{K2 campaigns}

\kepler~is poised to observe 10 fields in the next 2.5 years, which is the estimated time when the satellite runs out of fuel. As of this writing, only fields for campaigns 0 and 1 have been set; the remaining field locations are tentatively set but are subject to change if any scientific or engineering benefit from such a change is identified. Because of the engineering constraints discussed in the Introduction, the roll angle of the satellite needs to be fine-tuned to provide stable pointing. With any field change, the roll angle will vary substantially. Table \ref{tab:fields} summarizes current and tentative K2 campaigns. For the remainder of this study we consider the first 6 campaigns and provide details for the remaining campaigns online at {\tt http://keplerEBs.villanova.edu/K2}, where the information will be updated as soon as the campaign parameters are announced.

\begin{table}[t]
\caption{
\label{tab:fields}
Proposed K2 campaigns. At the time of this writing, campaigns 0 and 1 have been set and the remaining campaigns are subject to change, depending on the engineering requirements and community feedback.
}
\begin{center}
\begin{tabular}{cllll}
\hline \hline
Campaign: & End Date: & Center $\ra$ & Center $\dec$ & Comments: \\
\hline
0 & 2014-05-04 & $\hms{06}{46}{59.58}$ & $\dms{+21}{22}{47.1}$ & Near galactic anti-center, M35, NGC2304 \\
1 & 2014-07-23 & $\hms{11}{37}{55.65}$ & $\dms{+01}{11}{19.7}$ & North galactic cap \\
2 & 2014-10-14 & $\hms{16}{34}{43.63}$ & $\dms{-22}{48}{49.0}$ & Near galactic center, M4, M80, M19 \\
3 & 2015-01-05 & $\hms{22}{21}{06.01}$ & $\dms{-11}{36}{59.4}$ & South galactic cap, Neptune \\
4 & 2015-03-29 & $\hms{03}{45}{59.04}$ & $\dms{+18}{07}{49.7}$ & Pleiades (M45), Hyades, NGC1647 \\
5 & 2015-06-20 & $\hms{09}{19}{02.66}$ & $\dms{+14}{11}{41.0}$ & Beehive (M44), M67 \\
6 & 2015-09-11 & $\hms{14}{01}{11.20}$ & $\dms{-13}{16}{02.8}$ & North galactic cap \\
7 & 2015-12-03 & $\hms{19}{34}{16.22}$ & $\dms{-22}{38}{23.4}$ & Near galactic center, NGC6717 \\
8 & 2016-02-24 & $\hms{01}{04}{43.18}$ & $\dms{+05}{11}{52.2}$ & South galactic cap \\
9 & 2016-05-17 & $\hms{18}{23}{35.72}$ & $\dms{-24}{12}{12.8}$ & Galactic center, Baade's window, M8 \\
\hline
\end{tabular}
\end{center}
\end{table}

\begin{figure}[t!]
\begin{center}
\includegraphics[width=0.9\textwidth]{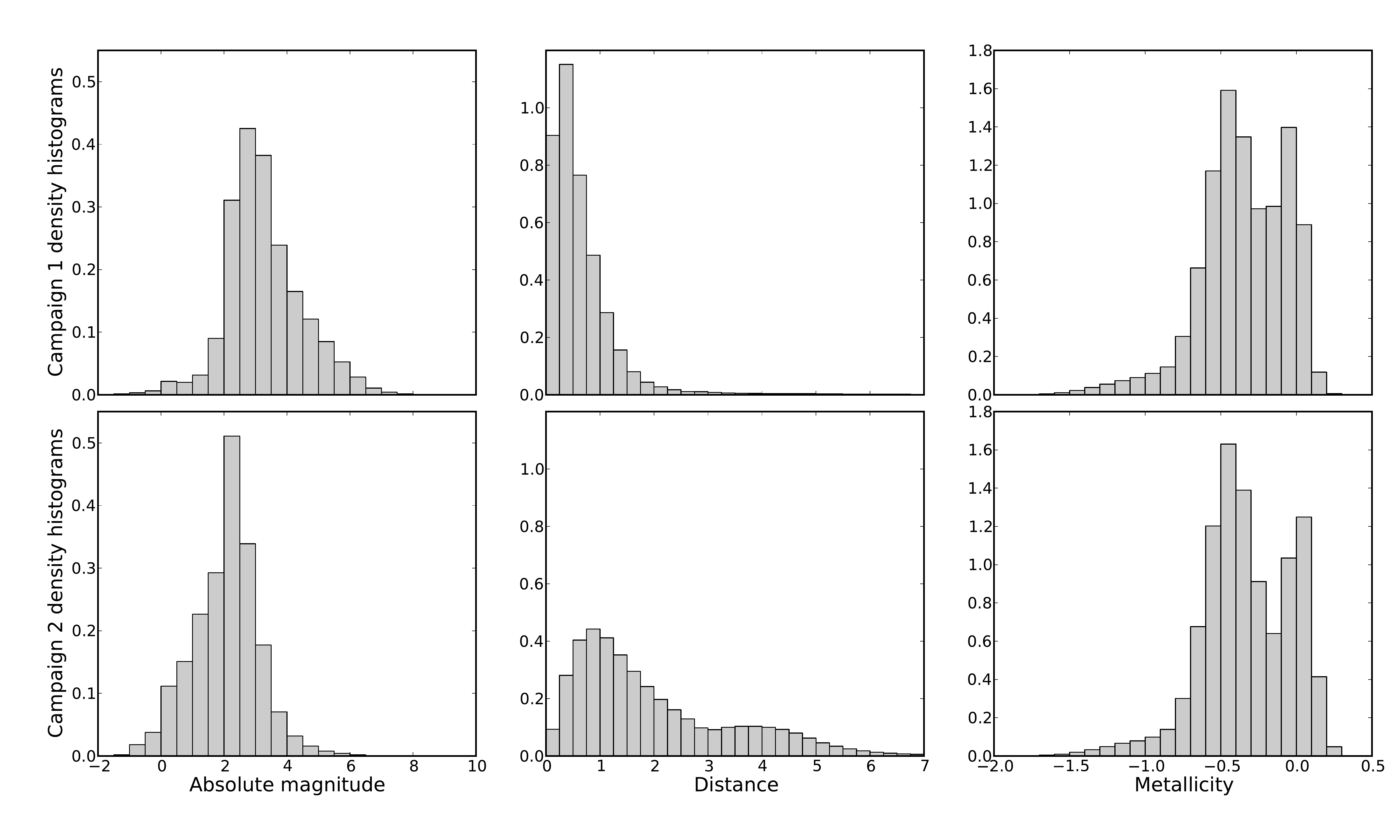} \\
\end{center}
\caption{
\label{fig:comparison}
Comparison of stellar populations for K2 campaigns 1 (north galactic cap) and 2 (galactic center) in absolute magnitude, distance and metallicity. The histograms are density-normalized to facilitate comparison. The field at higher galactic latitudes is more abundant in intrinsically fainter stars that are closer to us (because of the $Kp=$7-17 magnitude cut) and are predominantly underabundant. The field towards the galactic center is dominated by brighter stars that are farther from us (because of the bulge giants) and feature distinct populations -- underabundant in the disk and overabundant in the bulge. The main appeal of K2 is precisely this diversity in stellar populations.
}
\end{figure}

\begin{figure}[t]
\begin{center}
\includegraphics[width=\textwidth]{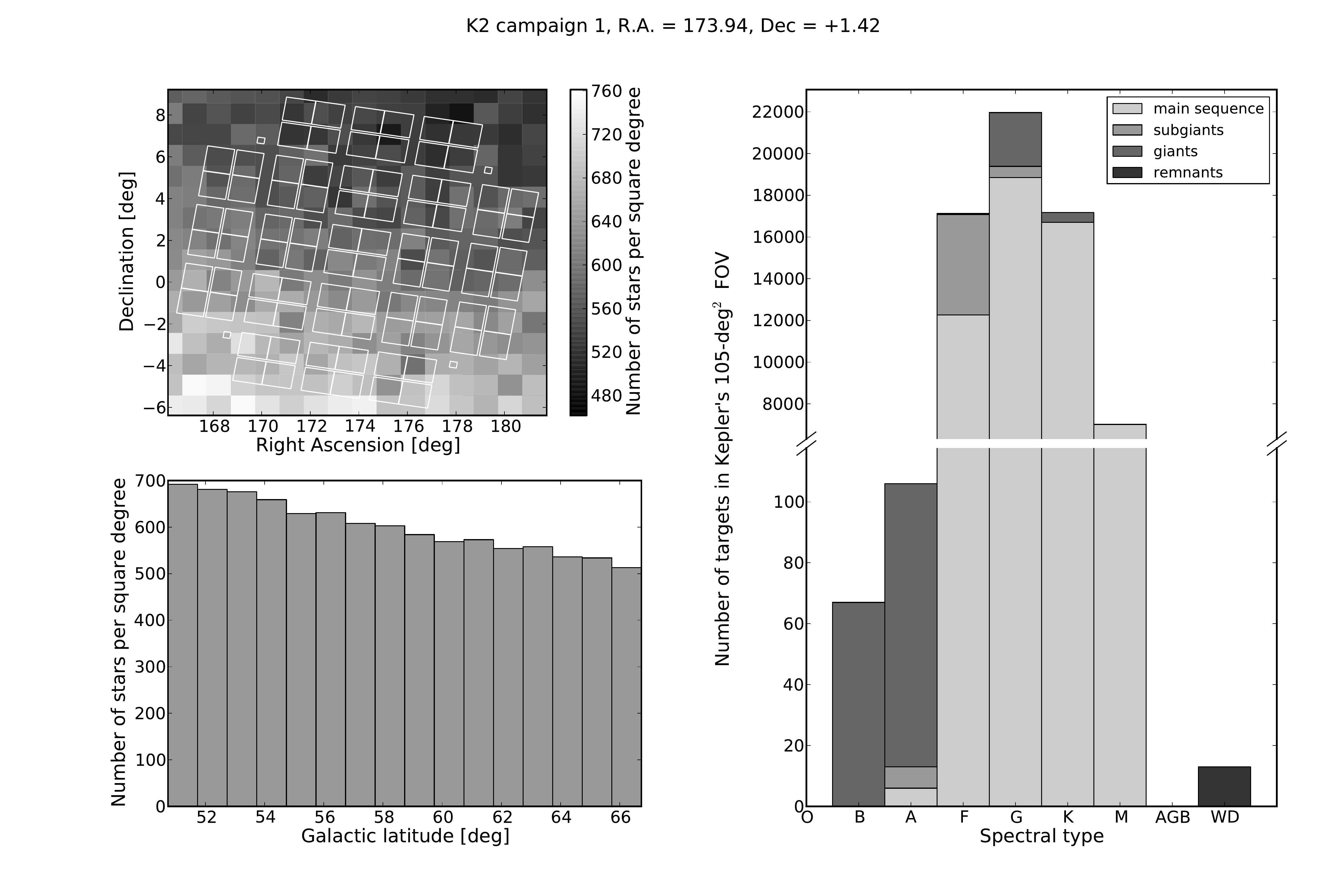} \\
\end{center}
\caption{
\label{fig:field1}
K2 campaign 1 ($\ra = \hms{11}{37}{55.65}$, $\dec = \dms{+01}{11}{19.7}$) statistics. Top left: the field-of-view superimposed on the star density map. This sparse field in terms of star counts features $\sim$500-700 stars per deg${}^2$. Bottom left: star count per deg${}^2$ as a function of galactic latitude. Right: star count as a function of spectral type, column-stacked by luminosity class. The sample consists of $Kp = $7-17 stars.
}
\end{figure}

Each campaign features a unique stellar population, with the most interesting targets listed in the Comments column of Table \ref{tab:fields}. We ran a detailed statistical check to estimate the expected field contents in terms of crowding and stellar populations. Fig.~\ref{fig:comparison} compares the populations in K2 campaigns 1 (north galactic cap) and 2 (galactic center) in the distributions in absolute magnitude, distance and metallicity. Fig.~\ref{fig:field1} provides an example for K2 campaign 1 that is scheduled to be observed in Summer 2014. The field is comparatively sparse in star counts brighter than $Kp \sim 17$, ranging between 500 and 700 per deg${}^2$. In comparison, the original \kepler~field featured over 7000 stars brighter than $Kp \sim 17$ per deg${}^2$. Statistics plots for the remaining campaigns are available online at the aforementioned Villanova \kepler~Eclipsing Binary Catalog site. Table \ref{tab:bulk} lists the number of stars of a given spectral type, including asymptotic giant branch (AGB) stars and white dwarfs (WD).

\begin{table}[t]
\caption{
\label{tab:bulk}
A compendium of stellar types in the first 6 K2 campaigns. AGB are asymptotic giant branch stars, WD are white dwarfs, and TP are transiting planets in 3 size ranges: 2-4\,$R_\oplus$, 4-8\,$R_\oplus$, and 8-32\,$R_\oplus$. The table is divided into main sequence stars (top) and subgiants, giants and remnants (bottom). Planet occurrence rates apply only to dwarf stars. Note that these numbers correspond to the entire fields at the $Kp = $7-17 range, not just selected targets.
}
\begin{center}
\tiny
\begin{tabular}{crrrrrrrrrrrr}
\hline \hline
\multicolumn{13}{c}{Main sequence stars} \\
\hline
Campaign: &   O &     B &      A &      F &      G &     K &  M & AGB & WD & TP${}_{2-4R_\oplus}$ & TP${}_{4-8R_\oplus}$ & TP${}_{8-32R_\oplus}$ \\
\hline
0 & 0 & 1640 & 16376 & 119056 & 177885 & 86731 & 11971 & 0 & 0 & $3464.5 \pm 213.2$ & $613.0 \pm 80.0$ & $346.5 \pm 53.3$ \\
1 & 0 & 0 & 6 & 12264 & 18852 & 16699 & 7006 & 0 & 0 & $316.6 \pm 19.5$ & $56.0 \pm 7.3$ & $31.7 \pm 4.9$ \\
2 & 0 & 26 & 523 & 206193 & 221858 & 105794 & 13113 & 0 & 0 & $4001.7 \pm 246.3$ & $708.0 \pm 92.3$ & $400.2 \pm 61.6$ \\
3 & 0 & 0 & 20 & 21643 & 29222 & 23344 & 8631 & 0 & 0 & $483.3 \pm 29.7$ & $85.5 \pm 11.2$ & $48.3 \pm 7.4$ \\
4 & 0 & 4 & 115 & 24175 & 45409 & 37481 & 9961 & 0 & 0 & $787.5 \pm 48.5$ & $139.3 \pm 18.2$ & $78.8 \pm 12.1$ \\
5 & 0 & 2 & 26 & 15627 & 26040 & 23806 & 8243 & 0 & 0 & $439.5 \pm 27.0$ & $77.8 \pm 10.1$ & $44.0 \pm 6.8$ \\
\hline
\end{tabular}

\bigskip

\begin{tabular}{crrrrrrrrr}
\hline \hline
\multicolumn{10}{c}{Subgiants, giants and remnants} \\
\hline
Campaign: &   O &     B &      A &      F &      G &     K &  M & AGB & WD \\
\hline
0 & 0 & 1250 & 15292 & 95791 & 54313 & 51664 & 1738 & 30 & 12 \\
1 & 0 & 67 & 100 & 4864 & 3114 & 471 & 6 & 0 & 13 \\
2 & 0 & 2534 & 4197 & 200750 & 330421 & 59162 & 613 & 18 & 11 \\
3 & 0 & 152 & 213 & 9568 & 6002 & 889 & 13 & 0 & 10 \\
4 & 0 & 30 & 119 & 11126 & 6681 & 2313 & 73 & 0 & 11 \\
5 & 0 & 50 & 81 & 6585 & 3512 & 853 & 20 & 0 & 7 \\
\hline
\end{tabular}

\normalsize
\end{center}
\end{table}

With \kepler~ultimately being a planet hunting mission, we can provide a rough estimate of the planetary occurrence rates as well. As part of their eta-Earth ($\eta_\oplus$) project, \citet{howard2010} observed a sample of 166 GK-type stars using the Keck/HIRES spectrograph and derived a power law that approximates the occurrence rates of close-in planets (orbital periods shorter than 50 days) as a function of planetary mass. This work was followed up for the \kepler~planetary candidate sample \citep{howard2012} to find occurrence rates of $13.0 \pm 0.8\%$ for 2-4\,$R_\oplus$ planets, $2.3 \pm 0.3\%$ for 4-8\,$R_\oplus$ planets, and $1.3 \pm 0.2\%$ for 8-38\,$R_\oplus$ planets. The numbers are in agreement with a study by \citet{fressin2013} that find a $16.5 \pm 3.6\%$ occurrence rate of 0.8-1.25\,$R_\oplus$ planets around main sequence FGK stars with orbital periods shorter than 85 days. They do not find any significant dependence of 0.8-4\,$R_\oplus$ planet occurrence rates on spectral type. While the majority of $\sim$2500 \kepler~planet candidates from the first 2 years of data have not been formally validated (even though the latest effort by \citet{rowe2014} added 340 planetary systems with 851 planets to the validated count), probabilistic simulations have shown that a vast majority of these candidates are in fact planets \citep{morton2011, fressin2013}. Orbital periods of exoplanets are consistent with a flat distribution in log space \citep{dong2013}, at least in the $\sim$1-day to the $\sim$100-day regime that is of most interest for K2 targets. We adopt these occurrence rates and the period distribution as definitive and use them to crudely estimate the number of transiting planets in each field.

For each star in the simulated field we draw an orbital period of a tentative planet from the flat $\log P$ distribution. Assuming that $M_\mathrm{planet} \ll M_*$, we compute the semi-major axis of the planetary orbit, assuming circular orbits. Orbital inclination is drawn from a uniform distribution and the orbit is projected onto the plane of sky. All systems that are sufficiently aligned with the line of sight to feature transits are counted, and their numbers are presented in Table \ref{tab:bulk}. These crude numbers do not account for the low S/N cutoff or single event systems, nor do they reflect any instrumental window functions. They only serve as a rough guide to the expected number of planets in the field around \emph{all} $Kp = 7$-17 stars, not only those selected as K2 targets.

\subsection{Crowding and contamination}

\kepler~is designed as a planet hunting mission, so it is crucial to understand and estimate the amount of crowding in the field and contamination due to third light. Eclipsing binary stars have been the main culprit for false positives: signals in light curves that resemble those of planetary transits \citep{fressin2013}. Because of third light dilution, the depths of stellar eclipses are quenched to planetary transit levels and complex approaches and/or follow-up spectroscopic campaigns are necessary to validate true planets \citep{torres2011}.

\begin{figure}[t]
\begin{center}
\includegraphics[width=0.65\textwidth]{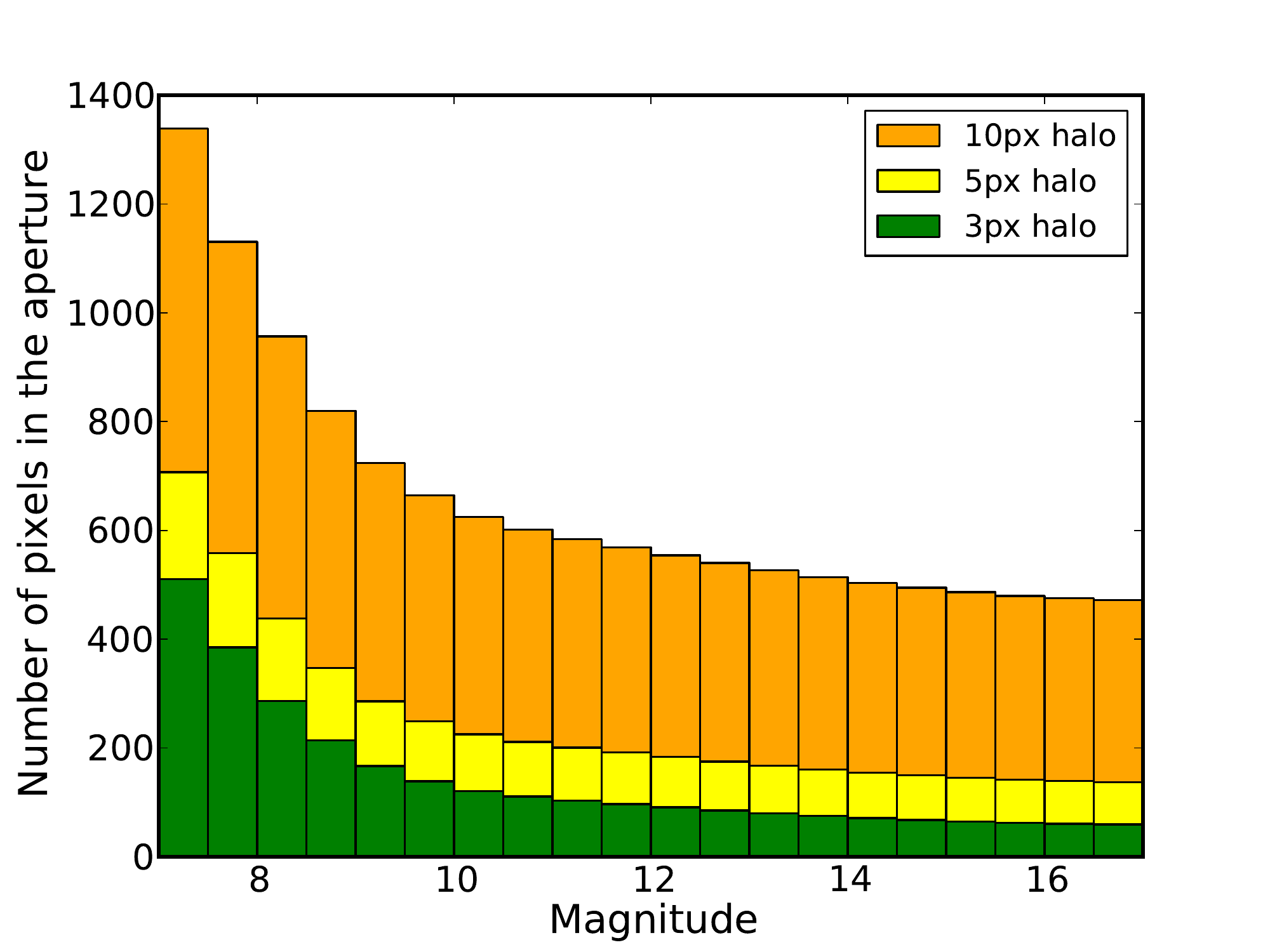} \\
\end{center}
\caption{
\label{fig:optapt}
Aperture size as a function of magnitude. Three cases are presented: optimal aperture size for the original \kepler~field augmented by a 10 pixel halo (orange), a 5 pixel halo (yellow) and a 3 pixel halo (green). A 10\,px halo is used for K2 campaign 0, 5\,px halo will likely be used for campaign 1, and a 3\,px halo might be used for subsequent campaigns, pending the results from the first two campaigns.
}
\end{figure}

Based on the engineering run in the ecliptic \citep{howell2014}, the K2 mission performs only marginally poorer than the original setup: the PSF is up to 5\% larger, and the main cause of photometric degradation is the solar-induced drift of the instrument. This drift is corrected by firing the on-board thrusters every 6 hours, so even though per-field PSF remains largely unchanged, the apertures are enlarged to account for this 1-2 pixel drift. The initial campaign serves primarily as an engineering run, and all  apertures are enlarged by adding a 10-pixel ``halo'' around them, to add ample margins for large smears and drift offsets. This will likely be reduced to a 5-pixel ``halo'' for campaign 1, and the aperture sizes will be further optimized for subsequent campaigns based on the engineering data. Fig.~\ref{fig:optapt} depicts new aperture sizes (in pixels) for a 10\,px halo, a 5\,px halo, and a 3\,px halo. This consideration, coupled with the decreased telemetry rate with the satellite that is currently $\sim$0.5\,au from Earth, the increased temporal baseline (and, with it, on-board data storage requirements) from 30 to 85 days between data downlinks, and a poorer data compression rate because of the target motion, requires a significant reduction in the number of targets observed: only 10,000 to 20,000 for campaigns 0 and 1, and $\sim$30,000 for the subsequent campaigns, compared to the 160,000 targets of the original mission.

With larger apertures, crowding becomes an even more important issue. As a crowding metric, we choose a $Kp=12$ magnitude star. The aperture size that corresponds to such a star is 554\,px, 184\,px and 92\,px for 10, 5 and 3\,px halos, respectively (cf.~Fig.~\ref{fig:optapt}). Given the \kepler~pixel size of $4'' \times 4''$, the corresponding areas are 2.46 arcmin${}^2$, 0.82 arcmin${}^2$ and 0.41 arcmin${}^2$. The star density per deg${}^2$ (Fig.~\ref{fig:density}) thus needs to be rescaled appropriately and we can use that as a metric for crowding in different fields. Table \ref{tab:crowding} lists these values for all K2 campaigns.

\begin{table}[t]
\caption{
\label{tab:crowding}
Crowding metric per K2 campaign. The values correspond to the (decimal) number of targets per $Kp=12$ star aperture with a 10\,px, 5\,px and a 3\,px halo. Minimal, average and maximal values per K2 campaign are provided. The values are subject to change with the field center and roll change.
}
\begin{center}
\scriptsize
\begin{tabular}{cccccccrrr}
\hline \hline
& \multicolumn{3}{c}{Min crowding} & \multicolumn{3}{c}{Average crowding} & \multicolumn{3}{c}{Max crowding} \\
Campaign: & 10\,px & 5\,px & 3\,px & 10\,px & 5\,px & 3\,px & 10\,px & 5\,px & 3\,px \\
\hline
0 & 2.159 & 0.720 & 0.360 & $4.049 \pm 1.198$ & $1.350 \pm 0.399$ & $0.675 \pm 0.200$ & 5.496 & 1.832 & 0.916 \\
1 & 0.351 & 0.117 & 0.058 & $0.410 \pm 0.039$ & $0.137 \pm 0.013$ & $0.068 \pm 0.006$ & 0.479 & 0.160 & 0.080 \\
2 & 3.247 & 1.082 & 0.541 & $7.513 \pm 3.785$ & $2.504 \pm 1.262$ & $1.252 \pm 0.631$ & 15.857 & 5.286 & 2.643 \\
3 & 0.508 & 0.169 & 0.085 & $0.658 \pm 0.107$ & $0.219 \pm 0.036$ & $0.110 \pm 0.018$ & 0.845 & 0.282 & 0.141 \\
4 & 0.590 & 0.197 & 0.098 & $0.891 \pm 0.227$ & $0.297 \pm 0.076$ & $0.148 \pm 0.038$ & 1.323 & 0.441 & 0.220 \\
5 & 0.420 & 0.140 & 0.070 & $0.553 \pm 0.091$ & $0.184 \pm 0.030$ & $0.092 \pm 0.015$ & 0.732 & 0.244 & 0.122 \\
\hline
\end{tabular}
\normalsize
\end{center}
\end{table}

\section{Conclusions}

The K2 mission concept promises to yield invaluable data similar in nature to the original \kepler~mission, and akin to the upcoming Transiting Exoplanet Survey Satellite (TESS; \citealt{ricker2010}). With $\sim$85 days on a single field, K2 will probe inherently different stellar populations and, contingent on NASA HQ approval and continued funding for 2.5 years, provide photometric coverage of over 250 thousand stars. This paper provides a guide to expected stellar populations, crowding and planetary occurrence rates along the ecliptic based on the improved Besan\c con model simulations.

Understanding stellar binarity and multiplicity is the next step in the study of K2 campaigns. From \kepler~observations of 2615 eclipsing and ellipsoidal binary stars \citep{prsa2011, slawson2011} that are essentially complete to $P \sim 500$ days, we can derive the underlying orbital period and eccentricity distributions. We do this by correcting for the bias using Bayesian methods outlined in \citet{hogg2010}. From the underlying distributions we simulate binary and multiple stars by applying the observed occurrence rates from \citet{raghavan2010} to the Besan\c con sample of stars that are grouped into multiple systems under the constraints of coevality and equal metallicity. These systems are then statistically examined for eclipses and eclipse timing variations. This in-depth analysis requires a subtantial discussion that is beyond the scope of the present paper.

Inherently different stellar (and planetary) populations along the ecliptic provide us with an opportunity to study population differences as a function of galactic latitude. Table \ref{tab:bulk} lists the expected numbers of main sequence stars and giants for each campaign, attesting to the variety of objects for which K2 will provide ultraprecise, long temporal baseline photometry. In combination with Gaia \citep{debruijne2012} that has recently seen first light, TESS that is scheduled to launch in 2017, and Plato \citep{catala2009} that has been selected as the third ETS medium-class science mission, the K2 dataset will be a gold mine for stellar and planetary astrophysics.

\acknowledgements

The authors acknowledge support through NASA Kepler PSP grant NNX12AD20G, and thank Kyle Conroy, Joshua Pepper, Jason Rowe, Tabetha Boyajian, Keivan Stassun, Pieter Degroote, Kelly Hambleton and William Borucki for useful discussions and suggestions.

\bibliography{k2pops}

\end{document}